\documentclass[final,5p,times,twocolumn,nopreprintline]{elsarticle}
\usepackage{amsmath,slashed,booktabs}
\usepackage{graphicx,graphics}
\usepackage{dcolumn}
\usepackage[hyperfootnotes=false]{hyperref}
\usepackage{xspace}
\usepackage{color}
\usepackage{balance}
\usepackage{multirow}

\usepackage{fancyhdr}
\fancypagestyle{firstpage}{%
	
	\lhead{}
	\rhead{\small INT-PUB-21-11, UWThPh 2021-3}
}
\newcommand{\GeV}{\,\text{GeV}}
\newcommand{\MeV}{\,\text{MeV}}
\newcommand{\keV}{\,\text{keV}}

\newcommand{\mpi}{M_\pi}
\newcommand{\mk}{M_K}

\renewcommand{\Im}{\text{Im}\,}
\newcommand{\Order}{\mathcal{O}}
\newcommand{\beq}{\begin{equation}}
\newcommand{\eeq}{\end{equation}}
\newcommand{\<}{\langle}
\renewcommand{\>}{\rangle}

\newcommand{\nn}{\nonumber\\}

\newcommand{\M}{\mathcal{M}}
\newcommand{\F}{\mathcal{F}}

\allowdisplaybreaks[1]

\begin{document}

\renewcommand{\theequation}{\arabic{equation}}

\begin{frontmatter}

\title{A dispersive estimate of scalar contributions to hadronic light-by-light scattering}   

\author[Mainz]{Igor Danilkin}
\author[Bern,Seattle]{Martin Hoferichter}
\author[Vienna,UCSD]{Peter Stoffer}

\address[Mainz]{Institut f\"ur Kernphysik and 
PRISMA$^+$ Cluster of Excellence, Johannes Gutenberg Universit\"at, 55099 Mainz, Germany}
\address[Bern]{Albert Einstein Center for Fundamental Physics, Institute for Theoretical Physics, University of Bern, Sidlerstrasse 5, 3012 Bern, Switzerland}
\address[Seattle]{Institute for Nuclear Theory, University of Washington, Seattle, WA 98195-1550, USA}
\address[Vienna]{University of Vienna, Faculty of Physics, Boltzmanngasse 5, 1090 Vienna, Austria}
\address[UCSD]{Department of Physics, University of California at San Diego, La Jolla, CA 92093, USA}

\begin{abstract}
  We consider the contribution of scalar resonances to hadronic
  light-by-light scattering in the anomalous magnetic moment of the muon.
  While the $f_0(500)$ has already been addressed in previous work using
  dispersion relations, heavier scalar resonances have only been estimated
  in hadronic models so far. Here, we compare an implementation of the
  $f_0(980)$ resonance in terms of the coupled-channel $S$-waves for
  $\gamma^*\gamma^*\to \pi\pi/\bar K K$ to a narrow-width approximation,
  which indicates $a_\mu^{\text{HLbL}}[f_0(980)]=-0.2(2)\times 10^{-11}$.
  With a similar estimate for the $a_0(980)$, the combined effect is thus
  well below $1\times 10^{-11}$ in absolute value. We also estimate the
  contribution of heavier scalar resonances. In view of the very uncertain situation concerning their
  two-photon couplings we suggest to treat them together with other
  resonances of similar mass when imposing the matching to short-distance
  constraints. Our final result is a refined estimate of the $S$-wave
  rescattering effects in the $\pi \pi$ and $\bar K K$ channel up to
  about $1.3$~GeV and including a narrow-width evaluation of the
  $a_0(980)$: $a_\mu^\text{HLbL}[\text{scalars}]=-9(1)\times 10^{-11}$.
\end{abstract}

\end{frontmatter}

\thispagestyle{firstpage}

\section{Introduction}

Hadronic light-by-light (HLbL) scattering currently gives the second-largest contribution to
the uncertainty in the Standard Model prediction for the anomalous magnetic moment of the muon~\cite{Aoyama:2020ynm,Aoyama:2012wk,Aoyama:2019ryr,Czarnecki:2002nt,Gnendiger:2013pva,Davier:2017zfy,Keshavarzi:2018mgv,Colangelo:2018mtw,Hoferichter:2019gzf,Davier:2019can,Keshavarzi:2019abf,Hoid:2020xjs,Kurz:2014wya,Melnikov:2003xd,Colangelo:2014dfa,Colangelo:2014pva,Colangelo:2015ama,Masjuan:2017tvw,Colangelo:2017qdm,Colangelo:2017fiz,Hoferichter:2018dmo,Hoferichter:2018kwz,Gerardin:2019vio,Bijnens:2019ghy,Colangelo:2019lpu,Colangelo:2019uex,Blum:2019ugy,Colangelo:2014qya}
\beq
\label{amuSM}
a_\mu^\text{SM}=116\,591\,810(43)\times 10^{-11}. 
\eeq
While at present the uncertainty of hadronic vacuum polarization dominates in the comparison to experiment~\cite{bennett:2006fi,Abi:2021gix,Albahri:2021ixb,Albahri:2021kmg,Albahri:2021mtf}
\beq
\label{exp}
a_\mu^\text{exp}=116\,592\,061(41)\times 10^{-11},
\eeq
see Refs.~\cite{Borsanyi:2020mff,Lehner:2020crt,Crivellin:2020zul,Keshavarzi:2020bfy,Malaescu:2020zuc,Colangelo:2020lcg} for recent developments in the comparison to lattice QCD, at the level of the final Fermilab precision, $\Delta a_\mu^\text{exp}=16\times 10^{-11}$, also the HLbL contribution needs to be improved. 
The phenomenological estimate from Ref.~\cite{Aoyama:2020ynm} (based on Refs.~\cite{Melnikov:2003xd,Colangelo:2014dfa,Colangelo:2014pva,Colangelo:2015ama,Masjuan:2017tvw,Colangelo:2017qdm,Colangelo:2017fiz,Hoferichter:2018dmo,Hoferichter:2018kwz,Gerardin:2019vio,Bijnens:2019ghy,Colangelo:2019lpu,Colangelo:2019uex,Pauk:2014rta,Danilkin:2016hnh,Jegerlehner:2017gek,Knecht:2018sci,Eichmann:2019bqf,Roig:2019reh})
\begin{equation}
\label{HLbL}
 a_\mu^\text{HLbL}=92(19)\times
10^{-11}
\end{equation}
agrees with $a_\mu^\text{HLbL}=82(35)\times 10^{-11}$ from lattice
QCD~\cite{Blum:2019ugy} (including the phenomenological estimate for the
charm contribution), and the average of both enters Eq.~\eqref{amuSM}. A
second, very recent lattice calculation~\cite{Chao:2021tvp} obtained
$a_\mu^\text{HLbL}=109.8(14.7)\times 10^{-11}$ (again after adding the
charm contribution), which agrees with both. 

The modern phenomenological approach to HLbL scattering is based on dispersion relations~\cite{Hoferichter:2013ama,Colangelo:2014dfa,Colangelo:2014pva,Pauk:2014rfa,Colangelo:2015ama}, to identify in a model-independent way the contributions from hadronic intermediate states. So far, the light pseudoscalar states $\pi^0$, $\eta$, $\eta'$ have been addressed in the dispersive approach, in which case then the uncertainty simply propagates from the transition form factors (TFFs) used as input~\cite{Masjuan:2017tvw,Hoferichter:2018dmo,Hoferichter:2018kwz,Gerardin:2019vio}. Similarly, the contributions from two-pion intermediate states have been evaluated in Refs.~\cite{Colangelo:2017qdm,Colangelo:2017fiz}, including rescattering effects in the $S$-wave, which arise as unitarization of the Born-term contributions and can be interpreted as a model-independent implementation of scalar resonances, in the case of two-pion intermediate states the $f_0(500)$. Here, we extend the analysis to a coupled-channel description including $\bar K K$ intermediate states, which allows us to study in more detail the energy region of the $f_0(980)$. In particular, we can then compare the full description in terms of partial-wave helicity amplitudes for $\gamma^*\gamma^*\to\pi\pi/\bar K K$~\cite{GarciaMartin:2010cw,Hoferichter:2011wk,Moussallam:2013una,Danilkin:2018qfn,Hoferichter:2019nlq,Danilkin:2019opj} with a narrow-width approximation (NWA).   

While the $\pi\pi/\bar K K$ system can be treated explicitly in terms of two-meson intermediate states, 
the viability of a description in terms of narrow resonances is important for estimates of higher-multiplicity contributions, making the $f_0(980)$ an interesting test case. Such estimates will be particularly important for axial-vector intermediate states, which play a special role~\cite{Melnikov:2003xd,Jegerlehner:2017gek,Roig:2019reh,Leutgeb:2019gbz,Cappiello:2019hwh,Masjuan:2020jsf} in the transition to short-distance constraints~\cite{Melnikov:2003xd,Bijnens:2019ghy,Colangelo:2019lpu,Colangelo:2019uex,Knecht:2020xyr,Ludtke:2020moa,Bijnens:2020xnl,Bijnens:2021jqo,Colangelo:2021nkr}.
The required TFFs are subject to a set of short-distance constraints themselves~\cite{Hoferichter:2020lap}, which combined with the available experimental input~\cite{Zanke:2021wiq} should provide enough information to reduce the part of the uncertainty in Eq.~\eqref{HLbL} attributed to axial-vector intermediate states and their interplay with short-distance constraints.  

However, existing estimates all rely on a simple Lagrangian definition of
such narrow-width (NW) contributions, which, in general, does not coincide
with a dispersive definition, and therefore cannot be combined with the
dispersive estimates for the one- and two-meson states. We will demonstrate
explicitly for the scalar case where the differences occur. In addition,
consistency of the dispersive approach requires a set of sum rules to be
fulfilled. For all single resonances other than pseudoscalar states these sum rules are in general not satisfied, which induces an ambiguity and renders
individual contributions dependent on the choice of the HLbL basis. Only the full result needs to fulfill the sum rules, restoring basis independence in the sum over all intermediate states.
For $S$-wave rescattering effects this potential ambiguity turns out to be
small, with the corresponding sum rule violated only at the level of $5\%$,
which allows us to provide an improved estimate of the scalar contributions
to HLbL scattering that by itself is essentially basis independent (and to be added to the pion- and kaon-box contributions). We also
comment on the role of even heavier scalar resonances and argue that those
should be included in the matching to short-distance constraints.

\section{Formalism}

\subsection{Hadronic light-by-light scattering}

We use the HLbL formalism established in Refs.~\cite{Colangelo:2017qdm,Colangelo:2017fiz}, and repeat here some of the salient features. First, following the general recipe by Bardeen, Tung~\cite{Bardeen:1969aw}, and Tarrach~\cite{Tarrach:1975tu}
(BTT), the HLbL tensor can be decomposed into $54$ Lorentz structures $T_i^{\mu\nu\lambda\sigma}$
\beq
\Pi^{\mu\nu\lambda\sigma} = \sum_{i=1}^{54} T_i^{\mu\nu\lambda\sigma} \Pi_i,
\eeq  
with scalar functions $\Pi_i$ that encode the dynamical content of the HLbL
amplitude. These $54$ $\Pi_i$, however, form a redundant set: the number of independent structures has to match the number of helicity amplitudes, in general $41$, whereof $27$ are of relevance in the $g-2$ case of one on-shell photon. A large portion of the formalism in Refs.~\cite{Colangelo:2017qdm,Colangelo:2017fiz} is thus devoted to defining a singly-virtual basis with $27$ elements $\check\Pi_i$, in terms of which the contribution from partial-wave helicity amplitudes can be analyzed. In order to formulate the result for HLbL scattering, it is useful to return to a linear combination of the original $\Pi_i$, denoted by $\hat \Pi_i$ in Refs.~\cite{Colangelo:2017qdm,Colangelo:2017fiz}, a subset $\bar \Pi_i$ of which contribute to the master formula
\begin{align}
\label{master}
a_\mu^\text{HLbL} &= \frac{\alpha^3}{432\pi^2} \int_0^\infty d\Sigma\,\Sigma^3 \int_0^1 dr\, r\sqrt{1-r^2} \int_0^{2\pi} d\phi \notag\\
&\times\sum_{i=1}^{12} T_i(Q_1,Q_2,Q_3) \bar\Pi_i(Q_1,Q_2,Q_3),
\end{align}
where the $T_i$ are known kernel functions and the Euclidean momenta squared are given
by~\cite{Eichmann:2015nra} 
\begin{align}
Q_{1}^2 &= \frac{\Sigma}{3} \left( 1 - \frac{r}{2} \cos\phi - \frac{r}{2}\sqrt{3} \sin\phi \right), \notag\\
Q_{1}^2 &= \frac{\Sigma}{3} \left( 1 - \frac{r}{2} \cos\phi + \frac{r}{2}\sqrt{3} \sin\phi \right), \notag\\
Q_3^2 &= \frac{\Sigma}{3} \left( 1 + r \cos\phi \right). 
\end{align}  
While the general result for higher partial waves becomes rather involved, the $S$-wave contribution can be written as~\cite{Colangelo:2017fiz}
\begin{align}
		\label{eq:SWaveDR}
		\hat\Pi_4^{J=0} &= \frac{1}{\pi} \int_{s_\text{thr}}^\infty ds' \frac{-2}{\lambda_{12}(s')(s'-q_3^2)^2} \Big( 4s' \Im h^0_{++,++}(s')\notag\\
		&- (s'+q_1^2-q_2^2)(s'-q_1^2+q_2^2) \Im h^0_{00,++}(s') \Big) , \notag\\
		\hat\Pi_{17}^{J=0} &= \frac{1}{\pi} \int_{s_\text{thr}}^\infty ds' \frac{4}{\lambda_{12}(s')(s'-q_3^2)^2} \Big( 2 \Im h^0_{++,++}(s')\notag\\
		&- (s'-q_1^2-q_2^2) \Im h^0_{00,++}(s') \Big) ,
\end{align}
where $\lambda_{12}(s)=\lambda(s,q_1^2,q_2^2)$, $\lambda(a,b,c)=a^2+b^2+c^2-2(ab+ac+bc)$ (plus crossed versions). In the special case of two-pion intermediate states, the threshold becomes $s_\text{thr}=4\mpi^2$ and the imaginary parts for the two-pion rescattering contribution for given isospin $I=0,2$ are
\begin{align}
\label{unitarity_pipi}
	\Im h^{J,I}_{\lambda_1\lambda_2,\lambda_3\lambda_4}(s) &= \frac{\sigma_\pi(s)\theta\big(s-4\mpi^2\big)}{32\pi} \Big[ h_{J,\lambda_1\lambda_2}^I(s) h_{J,\lambda_3\lambda_4}^{I*}(s)\notag\\
	&\qquad-c_I N_{J,\lambda_1\lambda_2}(s) N_{J,\lambda_3\lambda_4}(s)\Big],
\end{align}
subtracting the Born-term contributions
 involving $N_{J,\lambda_1\lambda_2}(s)$ and isospin factors $c_0=4/3$, $c_2=2/3$ (all amplitudes on the right-hand side refer to the $\gamma^*\gamma^*\to\pi\pi$ partial waves). The phase-space factor is $\sigma_\pi(s)=\sqrt{1-4\mpi^2/s}$. The subtraction of the Born terms in Eq.~\eqref{unitarity_pipi} is required to avoid a double counting of the pion-box contribution.
 
 The formulation as in Eq.~\eqref{eq:SWaveDR} corresponds to a particular
 choice of the singly-virtual basis $\check\Pi_i$, and the requirement that
 different bases be equivalent leads to a set of sum rules that need to be
 fulfilled (ultimately, as a consequence of the $T_i^{\mu\nu\lambda\sigma}$ not all having the same mass dimension). For the $S$-waves, there is only a single combination that
 contributes to these sum rules, amounting to a relation between integrals over the two possible $S$-wave helicity projections 
\begin{align}
		\label{eq:SR}
		0 &= \frac{1}{\pi} \int_{s_\text{thr}}^\infty ds' \frac{1}{\lambda_{12}(s')(s'-q_3^2)} \Big( 2 \Im h^0_{++,++}(s')\notag\\
		&- (s'-q_1^2-q_2^2) \Im h^0_{00,++}(s') \Big) +\text{higher waves}.
\end{align}
Basis independence requires that this sum rule be satisfied by the sum over all intermediate hadronic states. It is automatically fulfilled by the scalar QED amplitudes and thus the pion- and kaon-box contributions, but needs to be monitored when calculating, e.g., rescattering corrections~\cite{Colangelo:2017qdm,Colangelo:2017fiz}. For contributions that do not individually satisfy the sum rules, a basis change amounts to a reshuffling between contributions of different partial waves and/or hadronic intermediate states.

\subsection{Coupled-channel amplitudes for $\gamma^*\gamma^*\to\pi\pi/\bar K K$}

\begin{figure*}[t]
	\centering
	\includegraphics[width=0.4\linewidth,clip]{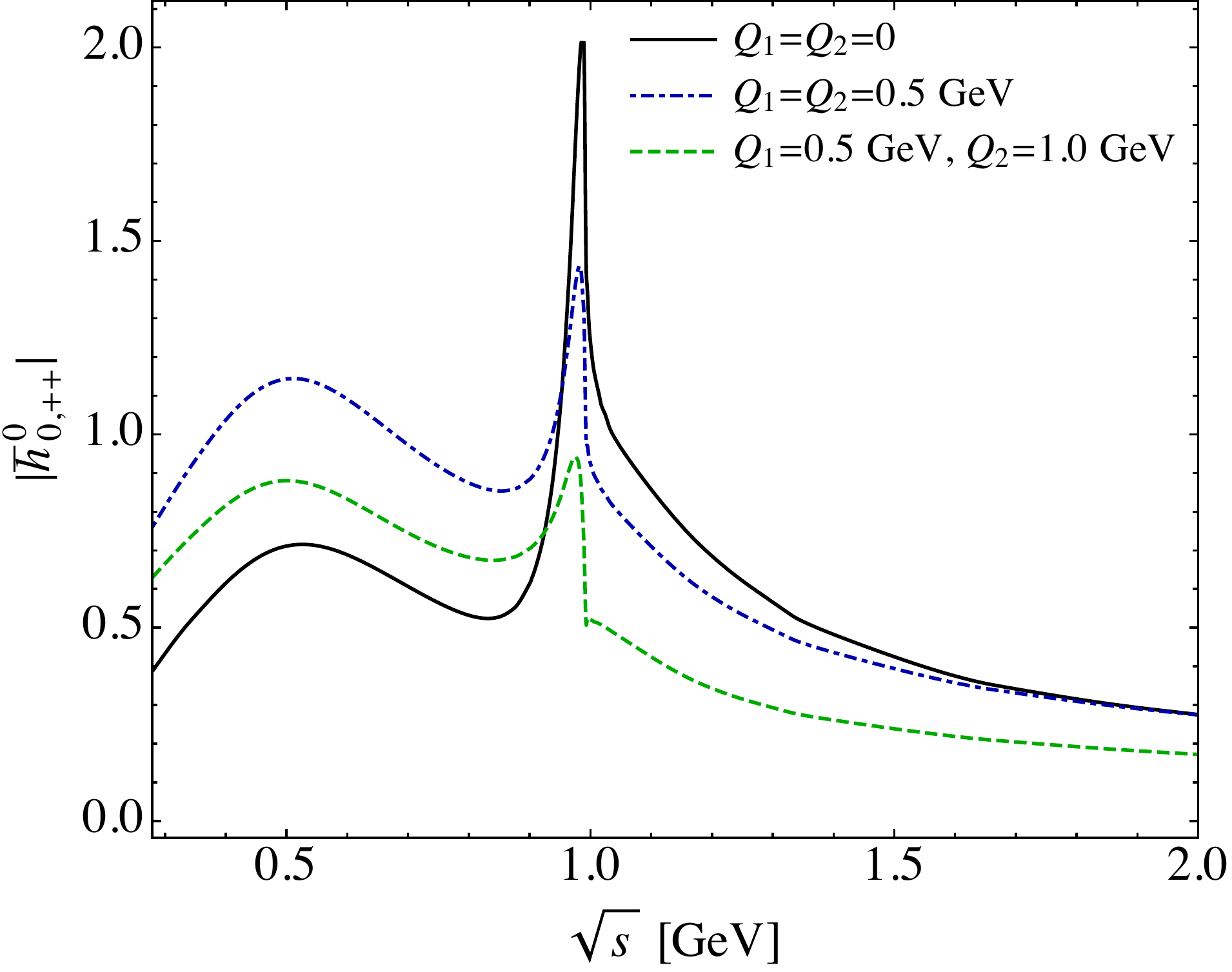}\qquad
	\includegraphics[width=0.4\linewidth,clip]{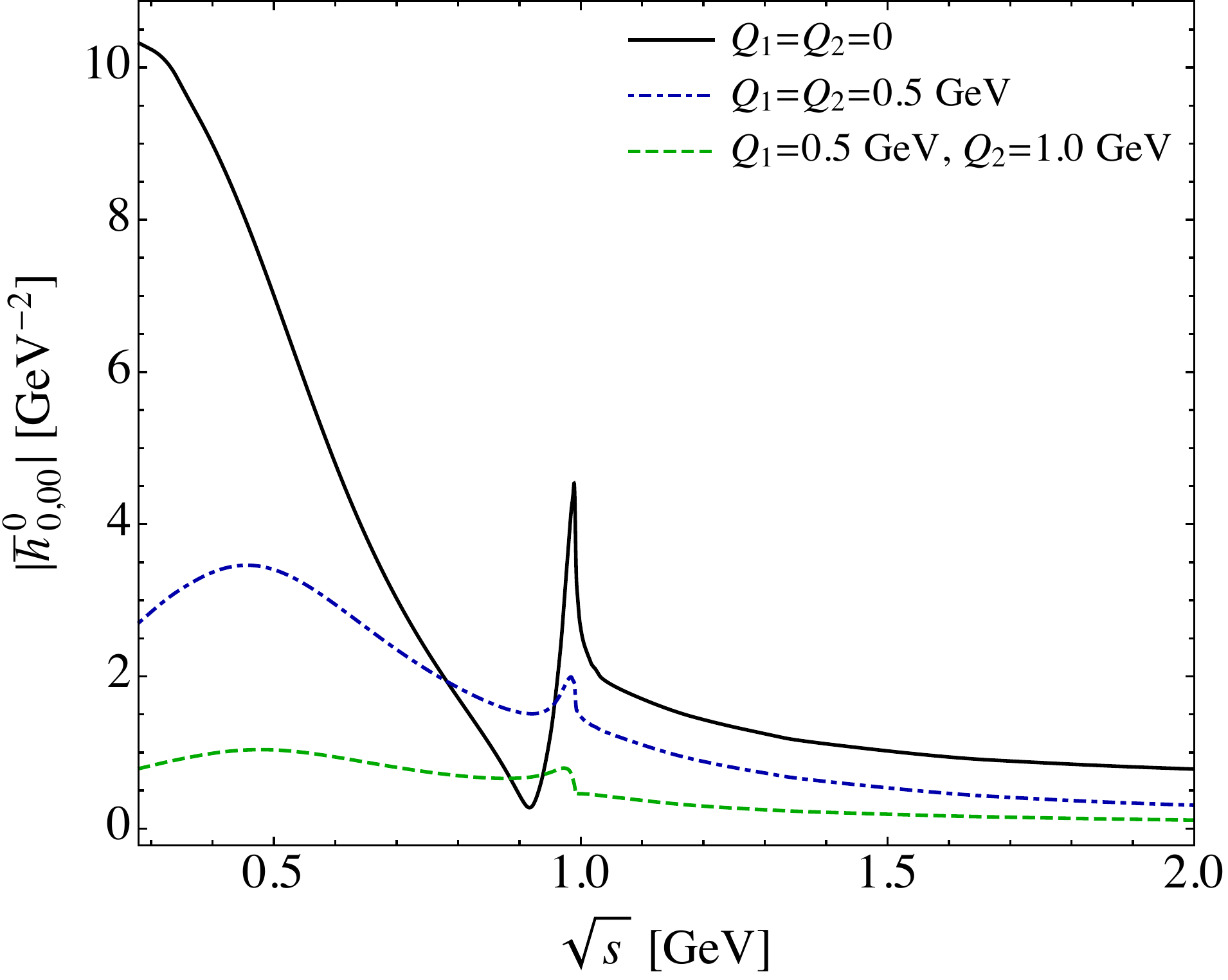}\\
	\includegraphics[width=0.4\linewidth,clip]{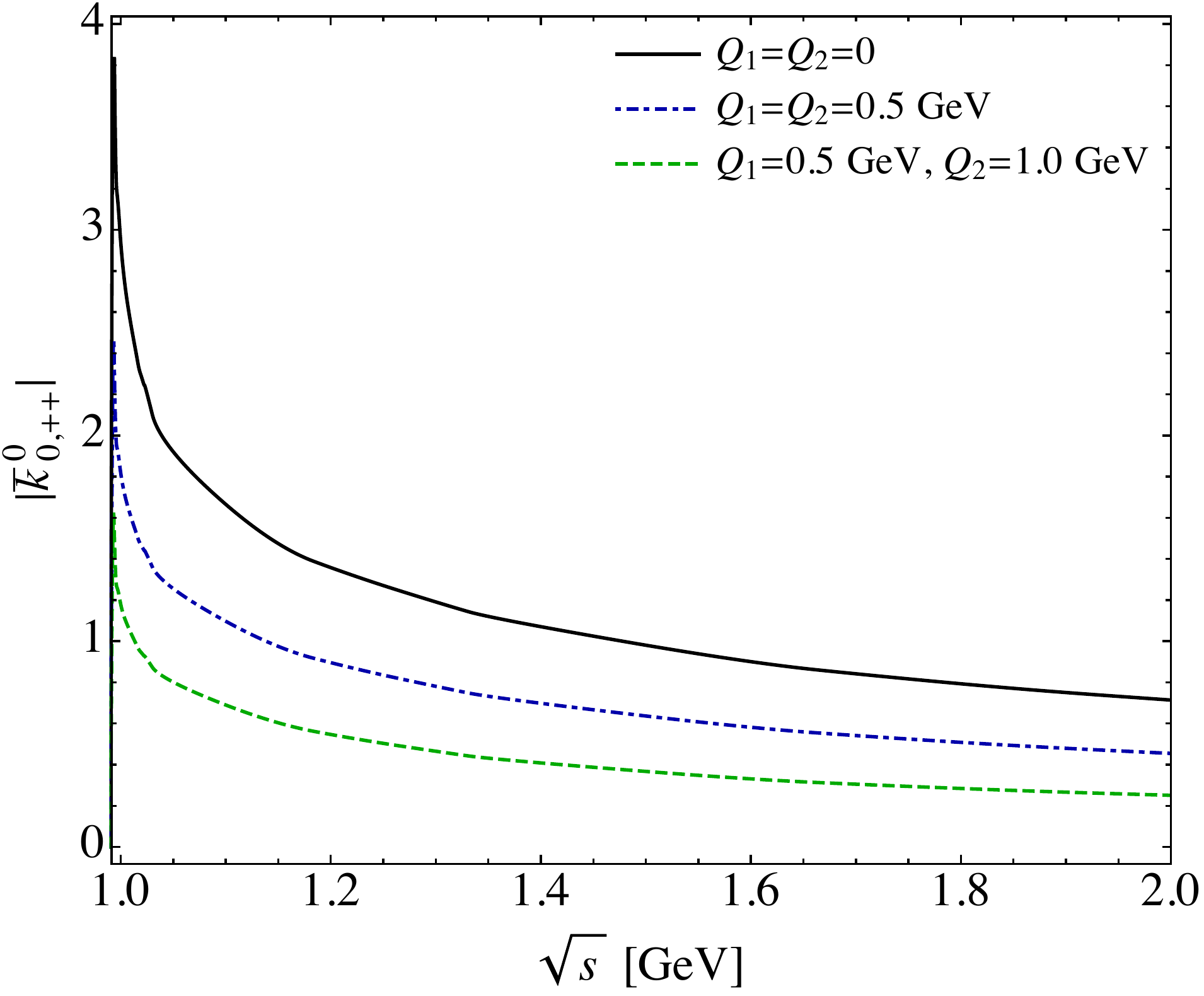}\qquad
	\includegraphics[width=0.4\linewidth,clip]{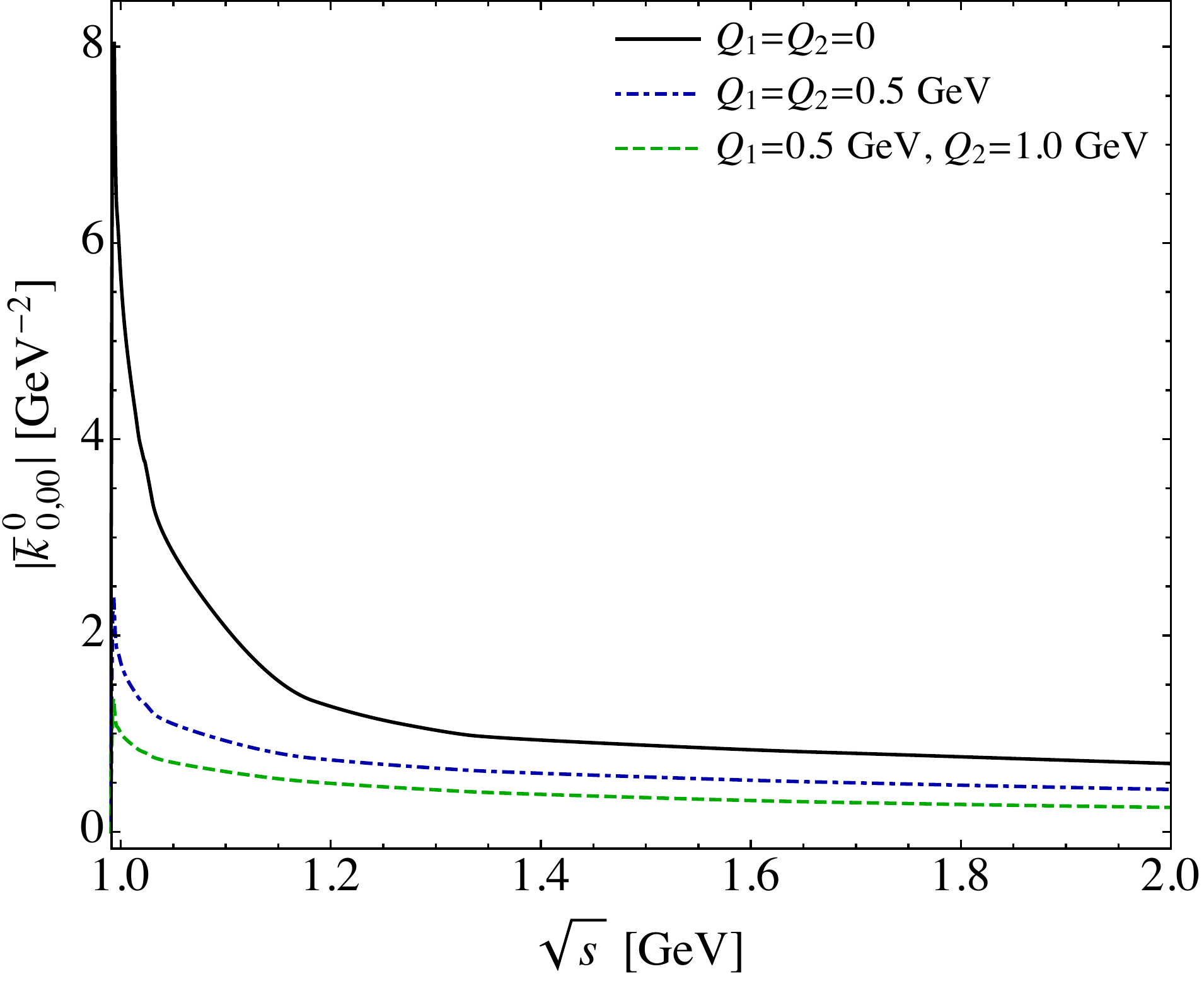}
	\caption{Modulus of the Born-term-subtracted partial waves $\bar h^0_{0,\lambda_1\lambda_2}$ and $\bar k^0_{0,\lambda_1\lambda_2}$, for a representative set of photon virtualities.}
	\label{fig:hk}
\end{figure*}

Including $\bar K K$ intermediate states, the unitarity relation~\eqref{unitarity_pipi} receives new contributions involving the full and Born-term amplitudes $k_{J,\lambda_1\lambda_2}$ and $M_{J,\lambda_1,\lambda_2}$, respectively. Note that these partial-wave amplitudes are normalized in such a way as to ensure the same unitarity condition for identical and non-identical particles~\cite{GarciaMartin:2010cw,Danilkin:2019opj}. Concentrating on the rescattering contribution for $I=J=0$, we have
\begin{align}
\label{unitarity_pipiKK}
	\Im h^{0,0}_{\lambda_1\lambda_2,\lambda_3\lambda_4}(s) &= \frac{\sigma_\pi(s)\theta\big(s-4\mpi^2\big)}{32\pi} \Big[ h_{0,\lambda_1\lambda_2}^0(s) h_{0,\lambda_3\lambda_4}^{0*}(s)\notag\\
	&\qquad-\frac{4}{3} N_{0,\lambda_1\lambda_2}(s) N_{0,\lambda_3\lambda_4}(s)\Big]\notag\\
	&+\frac{\sigma_K(s)\theta\big(s-4\mk^2\big)}{32\pi} \Big[ k_{0,\lambda_1\lambda_2}^0(s) k_{0,\lambda_3\lambda_4}^{0*}(s)\notag\\
	&\qquad-\frac{1}{2} M_{0,\lambda_1\lambda_2}(s) M_{0,\lambda_3\lambda_4}(s)\Big],
\end{align}
where the $\gamma^*\gamma^* \to \pi\pi$ and $\gamma^*\gamma^* \to K\bar{K}$ amplitudes are obtained using a modified coupled-channel Muskhelishvili--Omn\`es (MO) formalism~\cite{Danilkin:2019opj} (whose notation we follow). Besides eliminating all kinematic constraints in the partial-wave helicity amplitudes~\cite{Colangelo:2017fiz}, the MO formalism requires as input the knowledge of the left-hand cuts and the  hadronic Omn\`es functions~\cite{Omnes:1958hv}. The latter we take from a data-driven $N/D$ analysis~\cite{Danilkin:2020pak}, in which the fit is performed to the most recent Roy and Roy--Steiner results on $\pi\pi \to \pi\pi$~\cite{GarciaMartin:2011cn} and $\pi\pi \to \bar K K$~\cite{Pelaez:2020gnd}, respectively. By analytic continuation to the complex plane, this solution produces $f_0(500)$ and $f_0(980)$ poles at $\sqrt{s_{f_0(500)}}=458(10)^{+7}_{-15} - i\, 256(9)^{+5}_{-8}\MeV$ and $\sqrt{s_{f_0(980)}}=993(2)^{+2}_{-1} - i\,21(3)^{+2}_{-4}\MeV$, in good agreement with Refs.~\cite{Caprini:2005zr,GarciaMartin:2011jx,Moussallam:2011zg} (see Ref.~\cite{Danilkin:2020pak} for more details on the uncertainty estimates). Since the constructed $N/D$ solution is based on a once-subtracted dispersion relation, the obtained Omn\`es matrix is bounded  asymptotically. As for the left-hand cuts, it has been verified by comparison to the on-shell data from Refs.~\cite{Marsiske:1990hx,Boyer:1990vu,Behrend:1992hy,Mori:2007bu,Uehara:2008ep,Uehara:2009cka} that the MO formalism based on the pion- and kaon-pole left-hand cuts alone provides a good description of the $f_0(500)$ and $f_0(980)$ regions; see, e.g., Ref.~\cite{Aoyama:2020ynm}.\footnote{This statement no longer holds true for the $D$-waves, see Refs.~\cite{GarciaMartin:2010cw,Danilkin:2018qfn,Hoferichter:2019nlq,Danilkin:2019opj}, for which vector-meson left-hand cuts need to be included, as determined by the respective TFFs~\cite{Niecknig:2012sj,
Hoferichter:2012pm,Schneider:2012ez,Danilkin:2014cra,Hoferichter:2014vra,Hoferichter:2017ftn,Albaladejo:2020smb}.} In particular, the two-photon widths $\Gamma_{\gamma\gamma}[f_0(500)]=1.37(13)^{+0.09}_{-0.06}\keV$ and $\Gamma_{\gamma\gamma}[f_0(980)]=0.33(16)^{+0.04}_{-0.16}\keV$~\cite{Danilkin:2020pak}, come out consistent with other dispersive  extractions~\cite{ Hoferichter:2011wk, Moussallam:2011zg, Dai:2014zta}. For the pion and kaon electromagnetic form factors that enter in the pion- and kaon-pole contributions for virtual photons, we use the vector-meson-dominance (VMD)  expressions  
\begin{align}
 F_\pi^V(s)&=\frac{M_\rho^2}{M_\rho^2-s},\notag\\
 F_K^V(s)&=\frac{1}{2}\frac{M_\rho^2}{M_\rho^2-s}
 +\frac{1}{6} \frac{M_\omega^2}{M_\omega^2-s}
 +\frac{1}{3} \frac{M_\phi^2}{M_\phi^2-s},
\end{align}
given that the difference to the full description is negligible~\cite{Colangelo:2017fiz}. The resulting rescattering contributions to the pion and kaon partial waves are shown in Fig.~\ref{fig:hk} for a representative set of photon virtualities. Both the $f_0(500)$ and the $f_0(980)$ are clearly visible in $\bar h^0_{0,\lambda_1\lambda_2}$, while the impact of the $f_0(980)$ is also reflected by the threshold enhancement in $\bar k^0_{0,\lambda_1\lambda_2}$.

\subsection{Narrow-width approximation}
\label{sec:NWA}

For a narrow scalar resonance with mass $m_S$ we decompose the matrix element with two electromagnetic currents $j_\mathrm{em}^\mu$ according to~\cite{Hoferichter:2020lap}
\begin{align}
\M^{\mu\nu}(p \rightarrow q_1,q_2) &= i \int d^4x \, e^{i q_1 \cdot x} \< 0 | T \{ j_\mathrm{em}^\mu(x) j_\mathrm{em}^\nu(0) \} | S(p) \> \notag\\
&=\frac{\F_1^S(q_1^2,q_2^2)}{m_S} T_1^{\mu\nu}  + \frac{\F_2^S(q_1^2,q_2^2)}{m_S^3} T_2^{\mu\nu} ,
\end{align}
with Lorentz structures
\begin{align}
\label{BTT_scalar}
		T_1^{\mu\nu} &= q_1 \cdot q_2 g^{\mu\nu} - q_2^\mu q_1^\nu , \notag\\
		T_2^{\mu\nu} &= q_1^2 q_2^2 g^{\mu\nu} + q_1 \cdot q_2 q_1^\mu q_2^\nu - q_1^2 q_2^\mu q_2^\nu - q_2^2 q_1^\mu q_1^\nu,
\end{align}
and TFFs $\F_1^S$, $\F_2^S$. 
This BTT decomposition is again free of kinematic singularities, and the normalization of $\F_1^S$ can be related to the $S\to\gamma\gamma$ partial width $\Gamma_{\gamma\gamma}$:
\begin{equation}
	| \F_1^S(0,0) |^2 = \frac{4}{\pi \alpha^2 m_S} \Gamma_{\gamma\gamma} .
\end{equation}
The contribution to HLbL scattering follows most easily by the replacement
\begin{align}
 \Im h^0_{++,++}(s)&=\bigg(\frac{m_S^2-q_1^2-q_2^2}{2m_S}\F_1^S(q_1^2,q_2^2)+\frac{q_1^2q_2^2}{m_S^3}\F_2^S(q_1^2,q_2^2)\bigg)\notag\\
 &\times\frac{m_S^2-q_3^2}{2m_S}\F_1^S(q_3^2,0)\times \pi\delta(s-m_S^2),\notag\\
 \Im h^0_{00,++}(s)&=\bigg(\frac{1}{m_S}\F_1^S(q_1^2,q_2^2)+\frac{m_S^2-q_1^2-q_2^2}{2m_S^3}\F_2^S(q_1^2,q_2^2)\bigg)\notag\\
 &\times\frac{m_S^2-q_3^2}{2m_S}\F_1^S(q_3^2,0)\times \pi\delta(s-m_S^2),
\end{align}
leading to
\begin{align}
	\label{eq:hatFunctionsNarrowScalar}
 \hat\Pi_4&=\frac{\F_1^S(q_3^2,0)}{q_3^2-m_S^2}\bigg(\frac{\F_1^S(q_1^2,q_2^2)}{m_S^2}-\frac{m_S^2+q_1^2+q_2^2}{2m_S^4}\F_2^S(q_1^2,q_2^2)\bigg),\notag\\
 \hat\Pi_{17}&=\frac{\F_1^S(q_3^2,0)}{q_3^2-m_S^2} \frac{\F_2^S(q_1^2,q_2^2)}{m_S^4},
\end{align}
together with crossed versions for $t$- and $u$-channel exchange. We stress that in a Lagrangian model as formulated in Refs.~\cite{Pauk:2014rta,Knecht:2018sci}, the numerator of the prefactor of $\F_2^S$ in the bracket is altered to $m_S^2 + q_1^2 + q_2^2 \mapsto q_3^2 + q_1^2 + q_2^2$, i.e., while the residues of the scalar pole agree, the dispersive and the model description differ by non-pole terms. This difference could be removed by a further change in the HLbL basis, but then of course all other contributions to HLbL scattering would also need to be evaluated in this new basis, including the comparison to the rescattering corrections. 

In the NWA, the sum rule~\eqref{eq:SR} evaluates to
\beq
0=-\frac{1}{4m_S^4}\F_2^S(q_1^2,q_2^2) \F_1^S(q_3^2,0) + \text{other states},
\eeq
so that, unless $\F_2^S=0$, the contribution of a single resonance is not unique because the sum rules that ensure basis independence cannot be fulfilled by a narrow scalar alone. This statement applies to all resonances apart from pseudoscalar states, which are not affected by the sum rules. While a complete evaluation of HLbL must not depend on the choice of basis and fulfill the sum rules, this is not the case for individual intermediate states. In consequence, NW estimates for HLbL contributions necessitate the specification of the chosen HLbL basis, as a basis change amounts to a reshuffling of contributions from different intermediate states. Here, we take the basis from Refs.~\cite{Colangelo:2017qdm,Colangelo:2017fiz}, to contrast a NW description with one based on the $\gamma^*\gamma^*\to\pi\pi/\bar K K$ partial waves. 

As follows from Eq.~\eqref{BTT_scalar}, the second TFF $\F_2^S$ only contributes to doubly-virtual processes, so that no direct information from experiment exists. However,
for large virtualities the TFFs can be analyzed in a light-cone expansion, whose leading result gives~\cite{Hoferichter:2020lap}
\begin{align}
 \F_1^S(q_1^2,q_2^2)&=F_S^\text{eff}m_S\int_0^1 d u\,\frac{3(2u-1)^2\phi(u)}{u q_1^2+(1-u)q_2^2},\notag\\
 \F_2^S(q_1^2,q_2^2) 
 	&= F_S^\text{eff} m_S^3 \int_0^1 d u\, \frac{3u(1-u)\phi(u)}{( u q_1^2+(1-u)q_2^2)^2},
 \label{result_scalar}
\end{align}
with distribution amplitude $\phi(u)$ and an effective decay constant $F_S^\text{eff}$.
Inserting the asymptotic form $\phi(u)=6u(1-u)$, the two integrals become related, leading to the expansion
\begin{equation}
\label{LC_scalar}
 \F_1^S(q_1^2,q_2^2)=\frac{F_S^\text{eff} m_S}{Q^2}f^S(w),\qquad
 \F_2^S(q_1^2,q_2^2)=\frac{F_S^\text{eff} m_S^3}{Q^4}f^S(w),
 \end{equation}
with 
\beq
\label{defQ}
Q^2=\frac{q_1^2+q_2^2}{2},\qquad w=\frac{q_1^2-q_2^2}{q_1^2+q_2^2},
\eeq
and
\beq
f^S(w)=\frac{3}{2w^4}\bigg(3-2w^2+3\frac{1-w^2}{2w}\log\frac{1-w}{1+w}\bigg).
\eeq
These results can be contrasted with the quark model from Ref.~\cite{Schuler:1997yw}
\begin{align}
\label{TFF_Schuler}
  \frac{\F_1^S(q_1^2,q_2^2)}{\F_1^S(0,0)}\bigg|_{\text{\cite{Schuler:1997yw}}}&=\frac{m_S^2(3m_S^2-q_1^2-q_2^2)}{3(m_S^2-q_1^2-q_2^2)^2}= -\frac{m_S^2}{6Q^2}+\Order\Big(Q^{-4}\Big),\notag\\
 \frac{\F_2^S(q_1^2,q_2^2)}{\F_1^S(0,0)}\bigg|_{\text{\cite{Schuler:1997yw}}}&=-\frac{2m_S^4}{3(m_S^2-q_1^2-q_2^2)^2}=-\frac{m_S^4}{6Q^4}+\Order\Big(Q^{-6}\Big),
\end{align}
which thus reproduces the asymptotic relation $\F_2^S(q_1^2,q_2^2)/\F_1^S(q_1^2,q_2^2)=m_S^2/Q^2$ from Eq.~\eqref{LC_scalar}, while correctly interpolating to the normalization for $\F_1^S$ (to also implement the correct $w$ dependence more complicated TFF parameterizations would be necessary, see, e.g., Refs.~\cite{Hoferichter:2018kwz,Zanke:2021wiq}). We will use the prescription~\eqref{TFF_Schuler} to evaluate our NW estimates in Sec.~\ref{sec:NW}.

\section{Two-meson rescattering}

In Refs.~\cite{Colangelo:2017qdm,Colangelo:2017fiz} the effect of the $f_0(500)$ was estimated in terms of $S$-wave $\pi\pi$ rescattering to $a_\mu^\text{HLbL}[f_0(500)]=-9(1)\times 10^{-11}$, where the uncertainty was mainly attributed to the high-energy continuation of the amplitudes and, in practice, assessed by the violation of the sum rule that needs to be fulfilled to make the contribution basis independent. In particular, the $\pi\pi$ rescattering was implemented using the $\pi\pi$ phase shift from the inverse-amplitude method  in the Omn\`es solution for $\gamma^*\gamma^*\to\pi\pi$, to explicitly remove the effects of the $f_0(980)$ and ensure a smooth high-energy behavior. 

\begin{table}[t]
	\centering
	\begin{tabular}{l c r  r}
	\toprule
	$\Lambda\ [\text{GeV}]$&  & $0.89$ & $2.0$\\\midrule
	\multicolumn{2}{l}{pion (+ kaon) Born terms ($S$-waves)} & $-11.4$ & $-11.8$\\
	\multicolumn{2}{l}{$S$-wave $I=0$ rescattering} & $-10.0$ & $-9.8$\\\midrule
	\multirow{3}{4cm}{sum rule pion (+ kaon) \\ Born terms ($S$-waves)} & $++,++$ & $8.0$ & $8.4$\\
	& $00,++$ & $-9.2$ & $-9.6$\\
	& total & $-1.2$ & $-1.2$\\\midrule
	\multirow{3}{4cm}{sum rule\\ $S$-wave  $I=0$ rescattering} & $++,++$ & $6.9$ & $6.8$\\
	 & $00,++$ & $-7.3$& $-7.2$\\
	& total & $-0.4$ & $-0.4$\\
\bottomrule
	\end{tabular}
	\caption{$S$-wave Born-term and $I=0$ rescattering contributions (upper panel), and helicity components of the sum rule (lower panels), all in units of $10^{-11}$. $\Lambda^2$ denotes the cutoff in the $s$ integration.}
	\label{tab:rescattering}
\end{table}

For the coupled-channel solution from Ref.~\cite{Danilkin:2019opj} we find the results shown in Table~\ref{tab:rescattering}, very close to the estimate  
from Refs.~\cite{Colangelo:2017qdm,Colangelo:2017fiz}.\footnote{The Born terms correspond to the contributions that are subtracted in Eq.~\eqref{unitarity_pipiKK}. Resumming all Born-term partial waves would result in $\frac{3}{2}$ times the corresponding box contributions due to the different double-spectral regions of box and rescattering contributions~\cite{Colangelo:2017fiz}.} The numerical result is completely dominated by the two-pion contribution, but we verified that the imaginary part cancels between the $\pi\pi$ and $\bar K K$ channels, as required by the unitarity relation. We also checked the contribution to the sum rule~\eqref{eq:SR}, estimated by the difference when evaluated with the alternative basis from
Ref.~\cite{Colangelo:2014dfa}. As in Refs.~\cite{Colangelo:2017qdm,Colangelo:2017fiz} we observe a cancellation of the helicity components up to a remainder of $5\%$, indicating that sum-rule violations either due to the high-energy region in the $S$-waves or higher partial waves are indeed small. In fact, for the $S$-wave contributions of the pion and kaon Born terms the sum-rule violations are at least twice this relative size, but of course in this case the cancellation via the higher partial waves is exact. This implies that the $S$-waves can essentially be regarded as basis independent and the residual sum-rule violation be treated as an uncertainty estimate for the high-energy continuation. In combination with the $I=2$ estimates from Refs.~\cite{Colangelo:2017qdm,Colangelo:2017fiz}, $a_\mu^\text{HLbL}[\text{$S$-wave, $I=2$, $\Lambda=2\GeV$}]=1.1\times 10^{-11}$, we quote for the complete $S$-wave rescattering
\beq
\label{rescattering_final}
a_\mu^\text{HLbL}[\text{$S$-wave rescattering}]=-8.7(1.0)\times 10^{-11},
\eeq
where the uncertainty covers the high-energy region, sum-rule violations, and input quantities. 

\begin{figure*}[t]
	\centering
	\includegraphics[width=0.4\linewidth,clip]{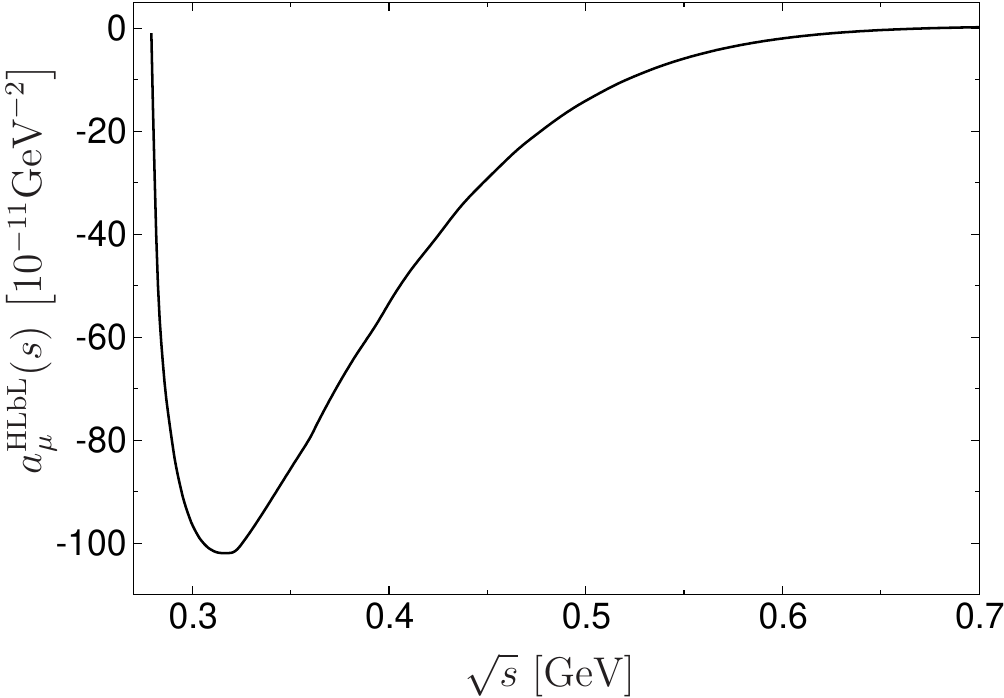}\qquad
	\includegraphics[width=0.4\linewidth,clip]{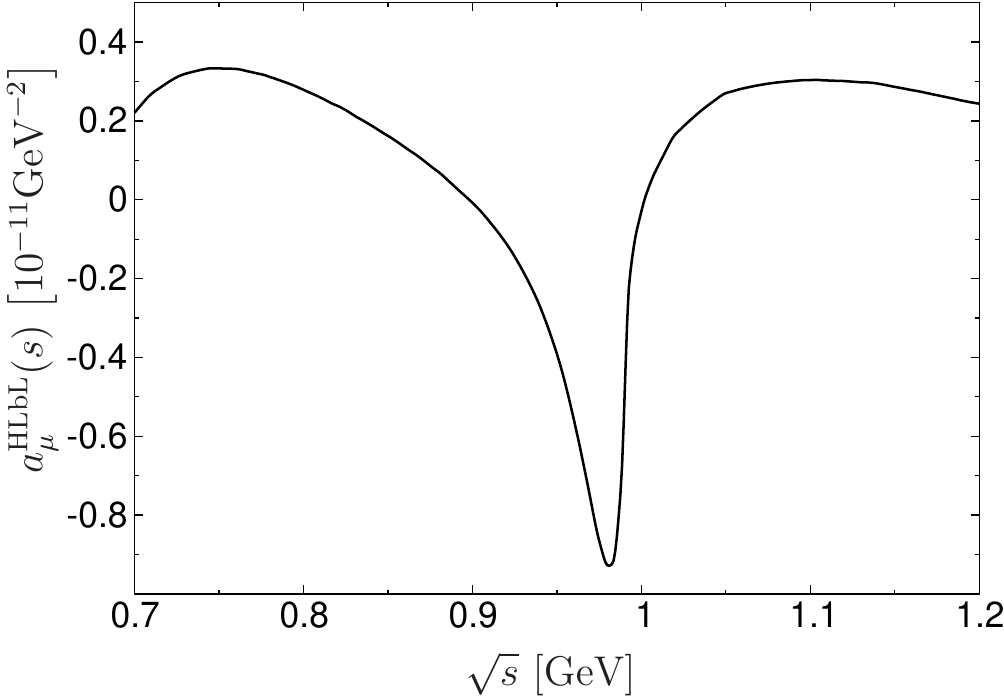}
	\caption{Integrand from Eq.~\eqref{amu_int} for the rescattering contribution, for the $f_0(500)$ (left) and the $f_0(980)$ (right). Note the different scales in both cases.}
	\label{fig:amus}
\end{figure*}

One could then define an $f_0(980)$ contribution by integrating  
over a window in $\sqrt{s}$, an obvious choice being
$\sqrt{s}\in[M_{f_0(980)}-\Gamma_{f_0(980)},M_{f_0(980)}+\Gamma_{f_0(980)}]$. The parameters quoted in Ref.~\cite{Zyla:2020zbs} are
\beq
M_{f_0(980)}=990(20)\MeV,\qquad \Gamma_{f_0(980)}=(10\text{--}100)\MeV,
\eeq
while dispersive analyses~\cite{GarciaMartin:2011jx,Moussallam:2011zg} favor a width around $50(20)\MeV$, which already reflects the complicated line shape of the $f_0(980)$ due to the close proximity of the $\bar K K$ threshold~\cite{Flatte:1976xu,Baru:2003qq}. 
We therefore suggest a different definition based on the decomposition 
\beq
\label{amu_int}
a_\mu^\text{HLbL}=\int_{4\mpi^2}^\infty ds' a_\mu^\text{HLbL}(s'),
\eeq
see Fig.~\ref{fig:amus} for the integrand produced by the rescattering corrections. This shows again that the $f_0(500)$ region by far dominates, while concentrating on the energy region around the $f_0(980)$ allows one to visualize the resulting line shape. Integrating the deficit below the baseline around $a_\mu^\text{HLbL}(s')=0.3\times 10^{-11}\GeV^{-2}$ gives $-0.18\times 10^{-11}$, which suggests an $f_0(980)$
contribution around
\beq
\label{rescatt}
a_\mu^{\text{HLbL}}[f_0(980)]\big|_\text{rescattering}= -0.2(1)\times 10^{-11}.
\eeq
The strong suppression compared to Eq.~\eqref{rescattering_final} happens to be of similar size as the suppression of the kaon to the pion box, $a_\mu^\text{HLbL}[\text{$\pi$-box}]/a_\mu^\text{HLbL}[\text{$K$-box}]\sim 32$~\cite{Aoyama:2020ynm}, which in this case is partly driven by $\mk^2/\mpi^2$, but given the complicated dependence of the kernel functions on the virtualities and the different line shapes the general mass scaling is difficult to anticipate.   
We stress that a contribution such as Eq.~\eqref{rescatt}, defined via a particular window in $\sqrt{s}$, is in general not basis independent, so that the comparison to the NWA described in the following section should refer to the same HLbL basis. 

\section{Narrow-width estimates}
\label{sec:NW}

The result from the rescattering approach can be contrasted with a description in terms of a narrow resonance. We use a dispersive definition in line with the HLbL basis chosen for the rescattering, and use the TFFs from the quark model of Ref.~\cite{Schuler:1997yw}, which provides a plausible interpolation between the normalization and the short-distance constraints derived in Ref.~\cite{Hoferichter:2020lap}. Using $M_{f_0(980)}=0.99\GeV$ and $\Gamma_{\gamma\gamma}[f_0(980)]=0.31(5)\keV$~\cite{Zyla:2020zbs}, we find 
\beq
\label{NW}
a_\mu^{\text{HLbL}}[f_0(980)]\big|_\text{NWA}=-0.37(6)\times 10^{-11},
\eeq
not too far away from the rescattering definition Eq.~\eqref{rescatt} (and an uncertainty referring only to $\Gamma_{\gamma\gamma}$). If the scale in the TFF parameterization~\eqref{TFF_Schuler} were lowered to a VMD expectation, $m_S\to M_\rho$, the NW result would move to $-0.26(4)\times 10^{-11}$, even closer to Eq.~\eqref{rescatt}.
Taken together with Eq.~\eqref{rescatt}, this would suggest the estimate
\beq
a_\mu^{\text{HLbL}}[f_0(980)]=-0.2(2)\times 10^{-11},
\eeq
and with $M_{a_0(980)}=0.98\GeV$ and $\Gamma_{\gamma\gamma}[a_0(980)]=0.3(1)\keV$~\cite{Zyla:2020zbs} a similar range would be expected for the $a_0(980)$. An improved evaluation of the isospin $I=1$ channel could be obtained from a coupled-channel analysis of the doubly-virtual helicity amplitudes for $\gamma^*\gamma^*\to\pi\eta/\bar KK$~\cite{Danilkin:2017lyn,Deineka:2018nuh,Lu:2020qeo}, following the same strategy as for the $f_0(980)$.
In fact, the coupled-channel analysis from Ref.~\cite{Lu:2020qeo} (based on the data from Refs.~\cite{Albrecht:1989re,Uehara:2009cf,Uehara:2013mbo}) prefers a width $\Gamma_{\gamma\gamma}[a_0(980)]=0.5^{+0.2}_{-0.1}\keV$, which translates to 
\beq
\label{a0_estimate}
a_\mu^{\text{HLbL}}[a_0(980)]=-\Big(0.6_{-0.1}^{+0.2}\Big)\times 10^{-11},
\eeq
or a slightly lower range, $a_\mu^{\text{HLbL}}[a_0(980)]=-\big(0.4_{-0.1}^{+0.2}\big)\times 10^{-11}$, if instead a VMD scale were used in the TFFs.     

We stress that when combining or comparing different contributions to HLbL, one should work within one unified framework, provided here by the dispersive framework and tensor basis of Refs.~\cite{Colangelo:2017qdm,Colangelo:2017fiz}. The basis dependence of the narrow scalar contribution can be illustrated by either setting $\F_2^S = 0$, or using the Lagrangian description by the replacement $m_S^2 + q_1^2 + q_2^2 \mapsto q_3^2 + q_1^2 + q_2^2$ as mentioned after Eq.~\eqref{eq:hatFunctionsNarrowScalar}, which with otherwise unchanged input for the TFFs (in particular, scale $m_S$) would result in
\begin{align}
	\label{eq:NoF2Lagrangian}
	a_\mu^{\text{HLbL}}[f_0(980)]\big|_\text{NWA}^{\F_2^S=0}=-0.47(8)\times 10^{-11} , \nn
	a_\mu^{\text{HLbL}}[f_0(980)]\big|_\text{NWA}^\text{Lagrangian}=-0.25(4)\times 10^{-11} .
\end{align}

In Ref.~\cite{Pauk:2014rta}, the $f_0(980)$ contribution is estimated in a Lagrangian model, keeping only the transverse helicity amplitude, which is then parameterized using a monopole form factor with scale varied between $(1\text{--}2)\GeV$, leading to a range $-(0.19\text{--}0.61)\times
10^{-11}$. The representation in terms of a single helicity amplitude combined with the Lagrangian definition resulted in kinematic singularities, which were removed by hand through angular averages. 
We emphasize that we cannot use the same input as Ref.~\cite{Pauk:2014rta} to reproduce these results using the BTT master formula for HLbL, in which a priori there are no kinematic singularities. A transverse $f_0(980)$ is obtained for $\F_2 = -2m_S^2/(m_S^2-q_1^2-q_2^2)\F_1$~\cite{Hoferichter:2020lap}, which with a monopole ansatz for $\F_1$ gives for the pole contribution a range $-(0.40\text{--}0.49)\times 10^{-11}$. Multiplying both form factors with an additional factor of $(m_S^2 - q_1^2 - q_2^2)/\lambda_{12}(m_S^2)$ would be closer in spirit to Ref.~\cite{Pauk:2014rta}, leading to a range $-(0.55\text{--}0.97)\times 10^{-11}$. Both variants are quite different from the range from Ref.~\cite{Pauk:2014rta} quoted above.

A NWA for the $f_0(980)$ is also considered in Ref.~\cite{Knecht:2018sci}, which uses the tensor decomposition~\eqref{BTT_scalar} without kinematic singularities, but again is based on a Lagrangian definition of the scalar contribution. The results are given as a function of a parameter $\kappa_S$, where $\kappa_S=0$ corresponds to switching off $\F_2^S$ and reduces $\F_1^S$ to a pure VMD form. The result without $\F_2^S$, $-0.42(9)\times 10^{-11}$, is close to Eq.~\eqref{eq:NoF2Lagrangian}, and quantifies the difference in the TFF input, where we believe that the quark model from Ref.~\cite{Schuler:1997yw} is more reliable because of the better implementation of the short-distance constraints~\cite{Hoferichter:2020lap} described in Sec.~\ref{sec:NWA} (see Refs.~\cite{Hoferichter:2020lap,Kroll:2016mbt} for the comparison to the singly-virtual data from Belle~\cite{Masuda:2015yoh}).
The difference to the results including $\F_2^S$, $-0.07(2)\times10^{-11}$, is mainly explained by their Lagrangian definition that includes non-pole pieces and to a lesser degree by the different TFF parameterizations. We checked that their spread for $\kappa_S \in [0,1]$ is much reduced when the dispersive basis of Refs.~\cite{Colangelo:2017qdm,Colangelo:2017fiz} is used instead.

References~\cite{Pauk:2014rta,Knecht:2018sci} also consider even heavier scalars, based on estimates of their two-photon coupling, e.g., $\Gamma_{\gamma\gamma}[f_0(1370)]=3.8(1.5)\keV$~\cite{Boglione:1998rw}. This estimate, however, describes a combined effect of $f_0(500)$ and $f_0(1370)$, which could not be reliably separated at the time. In more recent partial-wave analyses the $f_0(500)$ couplings can be isolated, while the effect of the $f_0(1370)$ is barely seen in $\gamma\gamma$ reactions. In fact, the number $\Gamma_{\gamma\gamma}[f_0(1370)]=4.0(1.9)\keV$ given in Ref.~\cite{Dai:2014zta} is accompanied by an explicit warning that even with its large error this number does not have the credibility of the other two-photon couplings (associating one star with the result). 
The situation is slightly better for the $a_0(1450)$, for which Ref.~\cite{Lu:2020qeo} quotes $\Gamma_{\gamma\gamma}[a_0(1450)]=1.05^{+0.50}_{-0.30}\keV$. Using U(3) assumptions, the decay widths of the excited scalars are related by
\begin{align}
\Gamma_{\gamma\gamma}[a_0(1450)]
&=\frac{\Gamma_{\gamma\gamma}[f_0(1370)]}{3\cos^2(\theta_A-\theta_0)}\frac{M_{a_0(1450)}}{M_{f_0(1370)}}\notag\\
&=\frac{\Gamma_{\gamma\gamma}[f_0(1500)]}{3\sin^2(\theta_A-\theta_0)}\frac{M_{a_0(1450)}}{M_{f_0(1500)}},
\end{align}
where $\theta_0=\arcsin (1/3)$ and $\theta_A$ is the mixing angle between $f_0(1370)$ and $f_0(1500)$~\cite{Zyla:2020zbs} (octet/singlet mixing is reproduced for $\theta_A=\pi/2$, see, e.g., Ref.~\cite{Zanke:2021wiq}). Since the $f_0(1500)$ has not been seen in $\gamma\gamma$ collisions~\cite{Zyla:2020zbs,Barate:1999ze,Acciarri:2000ex,Uehara:2008ep}, one could determine $\theta_A$ by the requirement that $\Gamma_{\gamma\gamma}[f_0(1500)]=0$. This choice, $\theta_A=\theta_0$, essentially defines an upper limit for $\Gamma_{\gamma\gamma}[f_0(1370)]\leq 3.0^{+1.4}_{-0.9}\keV$, where the uncertainties are propagated from Ref.~\cite{Lu:2020qeo} but do not include an additional U(3) uncertainty. Translating these couplings into a contribution to HLbL scattering is further complicated by the absence of information on the corresponding TFFs, and for the heavy scalars it makes a bigger difference if the scale is set by $m_S$ or VMD. The corresponding results are
\begin{align}
 a_\mu^{\text{HLbL}}[f_0(1370)]&=-(1.5^{+0.7}_{-0.4})\times 10^{-11}\quad \big[-(0.6^{+0.3}_{-0.2})\times 10^{-11}\big],\notag\\
 a_\mu^{\text{HLbL}}[a_0(1450)]&=-(0.5^{+0.2}_{-0.1})\times 10^{-11}\quad \big[-(0.2^{+0.1}_{-0.05})\times 10^{-11}\big],
\end{align}
where for the $f_0(1370)$ we adopted the above U(3) estimate and the
numbers in brackets are obtained for $m_S\to M_\rho$ in
Eq.~\eqref{TFF_Schuler}. The comparison for the $f_0(980)$ to the
implementation in terms of partial waves suggests that the latter results
may be more reliable, pointing to a combined effect of
$f_0(1370)$ and $a_0(1450)$ of at most $-1\times 10^{-11}$.  

However, given that even the two-photon couplings of the heavy scalar resonances are 
highly uncertain, let alone their TFFs, we propose a different point of view here. In comparison to the tensor mesons $f_2(1270)$ and $a_2(1320)$ these states are not seen prominently in two-photon reactions, and in contrast to axial-vector states they are not expected to play a special role in the implementation of short-distance constraints. Moreover, with a mass around $1.5\GeV$ and in view of the substantial uncertainties, it is not clear that a description in terms of hadronic degrees of freedom is useful, while it should be more promising to cover the respective physics in the asymptotic matching~\cite{Bijnens:2019ghy,Bijnens:2020xnl,Bijnens:2021jqo}.   
Comparing to the additional scalar contribution given in Ref.~\cite{Aoyama:2020ynm}, we have now included the $f_0(980)$ in the rescattering part~\eqref{rescattering_final}, and combined with the $a_0(980)$ NW estimate from Eq.~\eqref{a0_estimate} we quote
\beq
\label{scalars_final}
a_\mu^\text{HLbL}[\text{scalars}]=-9(1)\times 10^{-11}
\eeq
as our final result for the $S$-wave contribution to HLbL scattering to be added to the pion and kaon boxes.  
With the effects of even heavier states moved into the asymptotic matching, this eliminates the need for an additional scalar contribution.

\section{Summary and outlook}

In this paper we addressed scalar contributions to HLbL scattering in the framework of dispersion relations. First, we extended previous work on the $f_0(500)$ resonance, implemented via rescattering corrections to two-pion intermediate states, to the coupled-channel system of $\pi\pi/\bar K K$, which allowed us to identify a contribution from the $f_0(980)$ resonance. The result is reasonably close to an estimate using narrow resonances, given the uncertainties inherent in the scalar transition form factors. With a similar estimate for the $a_0(980)$ we find that the combined effect  
$a_\mu^{\text{HLbL}}[f_0(980)+a_0(980)]$ is well below $1\times 10^{-11}$,
so that our final result for the scalar contributions~\eqref{scalars_final}
is by far dominated by the $f_0(500)$ region. As for the contribution of
even heavier resonances, we argued that given scant
experimental input for their two-photon couplings and transition form
factors a description in terms of hadronic degrees of freedom is not
particularly useful and that their contribution should be included in the
asymptotic matching.   

Another important goal of this paper is to emphasize conceptual issues
that first arise for the scalar contributions, but will become more
critical for axial-vector and tensor resonances. First of all, it is only
in a dispersive framework that the narrow-width approximation corresponds
to including a pure pole term. Definitions based on phenomenological
Lagrangians usually include non-pole terms, which are model dependent,
modify the high-energy behavior, and have a significant numerical impact.
This observation is critical, because to ensure consistency of the entire
HLbL result, each contribution needs to be defined and evaluated
within the same framework. Even within the dispersive approach a basis
change for the HLbL tensor leads to an ambiguity in the definition of
each individual contribution, because a set of sum rules that guarantees basis independence
in general only needs to be satisfied by the sum over all hadronic intermediate states. We
demonstrated that for the $S$-wave rescattering the only relevant sum rule
is well fulfilled thanks to a cancellation between helicity components, so
that the result~\eqref{rescattering_final} is essentially basis
independent. For future estimates of axial-vector and tensor contributions
these subtleties will require a careful treatment, and together with the
matching to short-distance constraints will be critical to improve the
precision in the evaluation of HLbL scattering.

\section*{Acknowledgments}

We thank G.~Colangelo for very fruitful discussions and collaboration on parts of this work.
Financial support by the DOE (Grant Nos.\ 
DE-FG02-00ER41132 and DE-SC0009919) and the SNSF (Project No.\ PCEFP2\_181117) is gratefully acknowledged.  The work of I.D.\ was supported by the Deutsche Forschungsgemeinschaft (German Research Foundation) in part through the Collaborative Research Center (The Low-Energy Frontier of the Standard Model, Projektnummer 204404729 -- SFB 1044) and in part through the Cluster of Excellence (Precision Physics, Fundamental Interactions, and Structure of Matter, PRISMA$^+$ EXC 2118/1) within the German Excellence Strategy (Project No.\ 39083149).  
We thank the Institute for Nuclear Theory at the University of Washington for its kind hospitality and  the DOE for partial support during INT Workshop INT-19-74W, where this project was initiated. 
  
\bibliographystyle{apsrev4-1_mod}
\balance
\biboptions{sort&compress}
\bibliography{AMM}

\end{document}